\pdfoutput=1
\documentclass[acmtog]{acmart}
\author{Hsiao-yu Chen, Paul Kry, Etienne Vouga}

\usepackage{booktabs} 
\usepackage{bm}
\usepackage{enumitem}
\usepackage{adjustbox}

\setitemize{noitemsep,topsep=0pt,parsep=0pt,partopsep=0pt,leftmargin=*}
\setenumerate{noitemsep,topsep=0pt,parsep=0pt,partopsep=0pt,leftmargin=*}

\usepackage[tableposition = top]{caption}
\usepackage[ruled]{algorithm2e} 

\SetAlFnt{\small}
\SetAlCapFnt{\small}
\SetAlCapNameFnt{\small}
\SetAlCapHSkip{0pt}

\DeclareMathOperator*{\argmin}{arg\,min}

\newcommand{\bX}{\mathbf{X}}
\newcommand{\bY}{\mathbf{Y}}
\newcommand{\bW}{\mathbf{W}}

\acmJournal{TOG}

\hypersetup{draft}
\begin{document}
\title{Locking-free Simulation of Isometric Thin Plates}



\begin{abstract}
To efficiently simulate very thin, inextensible materials like cloth or paper, it is tempting to replace force-based thin-plate dynamics with hard isometry constraints. Unfortunately, naive formulations of the constraints induce membrane locking---artificial stiffening of bending modes due to the inability of discrete kinematics to reproduce exact isometries. We propose a simple set of meshless isometry constraints, based on moving-least-squares averaging of the strain tensor, which do not lock, and which can be easily incorporated into standard constrained Lagrangian dynamics integration.
\end{abstract}

%
%


%
%


\begin{teaserfigure}
\centering
\includegraphics[width=\textwidth]{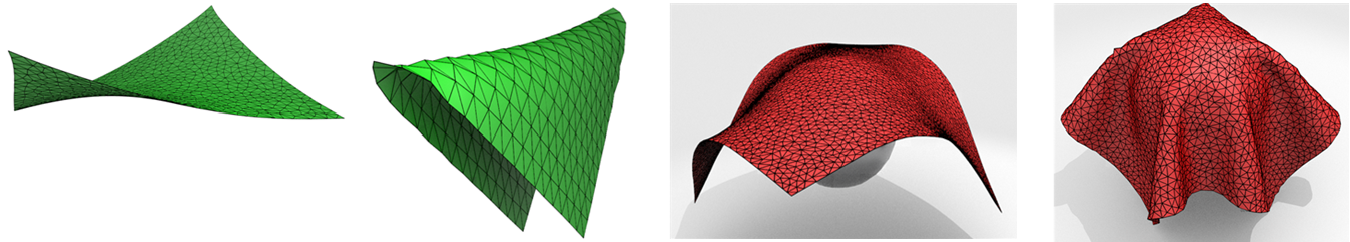}
\caption{Standard force-based simulations of thin plates severely membrane-lock when the thickness is small (left on each pair). Our method replaces membrane dynamics with simple isometry constraints that don't lock (right).}
\label{fig:teaser}
\label{fig:locking}
\end{teaserfigure}
\maketitle

\section{Introduction}
Efficiently simulating thin objects such as cloth, paper, or skin remains significantly more challenging than simulating their rigid or volumetric counterparts, due to their kinematic complexity and nonlinear physics. Specifically, numerical discretizations of thin plates must overcome two challenges: the dramatic stiffness disparity between stretching and bending material forces, and the tendency for discretizations of thin surfaces to \emph{lock}. 
\begin{itemize}
    \item \textbf{Stiffness scale-separation.} The behavior of thin plates is governed by a combination of \emph{membrane} forces, resisting in-plane stretching and shear of the material, and \emph{bending} forces. These forces have dramatically different stiffnesses: stretching stiffness scales like the thickness of the material, whereas bending scales like thickness cubed; cloth, for example, is approximately $0.25\,\mathrm{mm}$ thick, so that the stretching forces are $16$ million times stronger. 
    
    For sufficiently thin objects, the scale separation between these stiffnesses is so vast that stretching of the material is imperceptible; in this setting paying the price in element or time step size in order to resolve the stretching modes does not make sense. Instead, one can look at \emph{isometric} kinematics, where constraints enforcing zero in-plane strain replace force-level resistance to stretching. However, such a strategy requires discrete surface kinematics that supports cleanly decoupling bending from stretching modes.
    \item \textbf{Membrane locking.} A smooth surface can bend into a general developable surface isometrically; the same is not true for a triangle mesh. Regular equilateral meshes can isometrically bend about their three axes of symmetry only; irregular meshes cannot bend at all without distorting some of the triangle edges. As a consequence, any formulation of stretching forces based on triangle edge lengths will suffer from \emph{membrane locking}: situations, such as holding a piece of cloth up by two corners as shown in Figure~\ref{fig:locking}, left, where deformations that are bending-dominated in the continuous regime are stretching-dominated in the discrete regime~\cite{Quaglino2016}, due to failure of bending to kinematically decouple from stretching. Note that unlike other discretization errors, locking is \emph{independent} of mesh resolution: an irregular mesh with inextensible edges will not bend no matter how fine.
    
    On the other side of the same coin, insufficiently constraining discrete inextensible plates, so that stretching-dominated deformations in the continuous regime are instead bending-dominated (or completely uncontrolled), allows equally undesirable \emph{spurious deformation modes}.
\end{itemize}

\paragraph{Main Idea} We present a method for simulating isometric thin plates by combining two simple ideas: replacing the very stiff membrane forces by hard constraints, and doing so on a formulation of the strain tensor that is averaged over surface patches via moving least squares, to avoiding membrane locking. We can then simulate the material by adding a bending energy and enforcing the hard inextensibility constraints using standard methods for constrained Lagrangian dynamics, such as the method of Fast Projections~\cite{GHFBG07}.

The resulting numerical method is simple to implement, and because stretching forces are replaced with hard constraints, it can simulate infinitesimally thin, inextensible materials without requiring very fine meshes or small time steps. Crucially, the averaged inextensibility constraints do not lock, and do not exhibit spurious modes, so that even very coarse, efficient simulations of thin materials undergoing large amounts of bending and crumpling give good qualitative results.
\section{Related Work}
There is a deep body of work in both computer graphics and finite element analysis on the simulation of thin plates and shells. Membrane locking is a well-known phenomenon for simulations involving low-order elements, and has been studied extensively by Quaglino~\shortcite{QuaglinoPhD,Quaglino2016}, who proposed several tests characterizing locking and taxonomized the locking behavior of a variety of triangle-mesh-based kinematics.  English and Bridson~\shortcite{English2008} suggested gluing triangles at edge midpoints rather than at vertices, which avoids locking but unfortunately suffers from spurious modes. Popular workarounds to locking include adaptive refinement of the mesh, as in the popular ArcSim~\cite{Narain2012} code and its extensions; avoiding triangles entirely and using higher-order elements; or compensating for locking by tweaking stiffness parameters on an ad-hoc per-example basis.

\paragraph{Isometry/Strain Limiting} Treating membrane strain with constraints rather than forces has proven to be a powerful technique for reproducing characteristic wrinkle patterns in thin shell materials. Early application of this idea was used to introduce fine wrinkles where contact introduces compression of the materials~\cite{Provot95,Bridson:2003:SCF:846276.846281}. Goldenthal et al.~\shortcite{GHFBG07} proposed a framework for constraint-based limiting of strain in the warp and weft directions on a quadrilateral mesh, and demonstrated that constraint-based inextensibility is significantly more efficient than force-based simulation of membrane strain, even when using implicit methods. Chen and Tang~\shortcite{Chen2010} enforce isometry in a least-squares sense while also respecting collision constraints. Other methods for enforcing strain limits have been proposed, with support for constraining all elements of the strain tensor~\cite{Thomaszewski2009} and for boosting performance by applying constraints in a hierarchical manner~\cite{Wang:2010:MRI}, though these methods treat strain based on triangle deformations and will lock unless the material is sufficiently compliant when under the strain limit. Also related is the method of position-based dynamics~\cite{MULLER2007109,Stumpp2008}, a stable and fast alternative to traditional physical simulation that replaces \emph{all} forces (even the soft bending forces) with constraints.

\paragraph{Tension Field Theory} In tension-dominated problems, wrinkles perpendicular to the direction of tension have no effect on the coarse material shape, and so can be neglected; this idea is at the heart of tension field theory~\cite{mansfield64,Steigmann90}, which has been discretized in graphics for predicting the inflated shape of balloons~\cite{Skouras2014}. The theory has also been applied to cloth, either via inequality constraints on edge lengths~\cite{Jin2017} or incorporation of wrinkle parameters into the kinematics~\cite{Quaglino2016}.

\paragraph{Fast Projections} We adopt Goldenthal et al.~\shortcite{GHFBG07}'s popular~\cite{dinev2018FEPR} observation that constrained Lagrangian dynamics can be substantially accelerated by relaxing the projection step. Modifications exist that improve convergence, especially for conflicting constraints~\cite{Bouaziz2014ProjectiveDF}.

\paragraph{Meshless Methods} Although most cloth solvers are mesh-based, meshless methods have also been explored~\cite{Yuan2008}. They are particularly appealing for problems involving fracture; peridynamics~\cite{Silling2000} has had significant success in computer graphics~\cite{Levine2015,Chen2018,He2018,Xu2018} and has been extended to shells~\cite{CHOWDHURY2016110}. Similar in spirit are Elastons~\cite{Martin2010}, a quadrature scheme for deformation energy based on unstructured sample points which can be used to simulate elastic bodies of arbitrary codimension in a unified way. Also related are the ``F-bar'' methods in finite elements for simulating incompressible volumes, which avoids locking by imposing area/volume constraints on patches of elements rather than on single quadrilaterals~\cite{deSouzaNeto1996} or triangles~\cite{Neto2005}. This idea has been applied to simulation of incompressible volumes~\cite{Ortiz2015,Boroomand2004}, including in graphics~\cite{Irving2007}.

\begin{figure*}[ht]
    \includegraphics[width=0.24\textwidth]{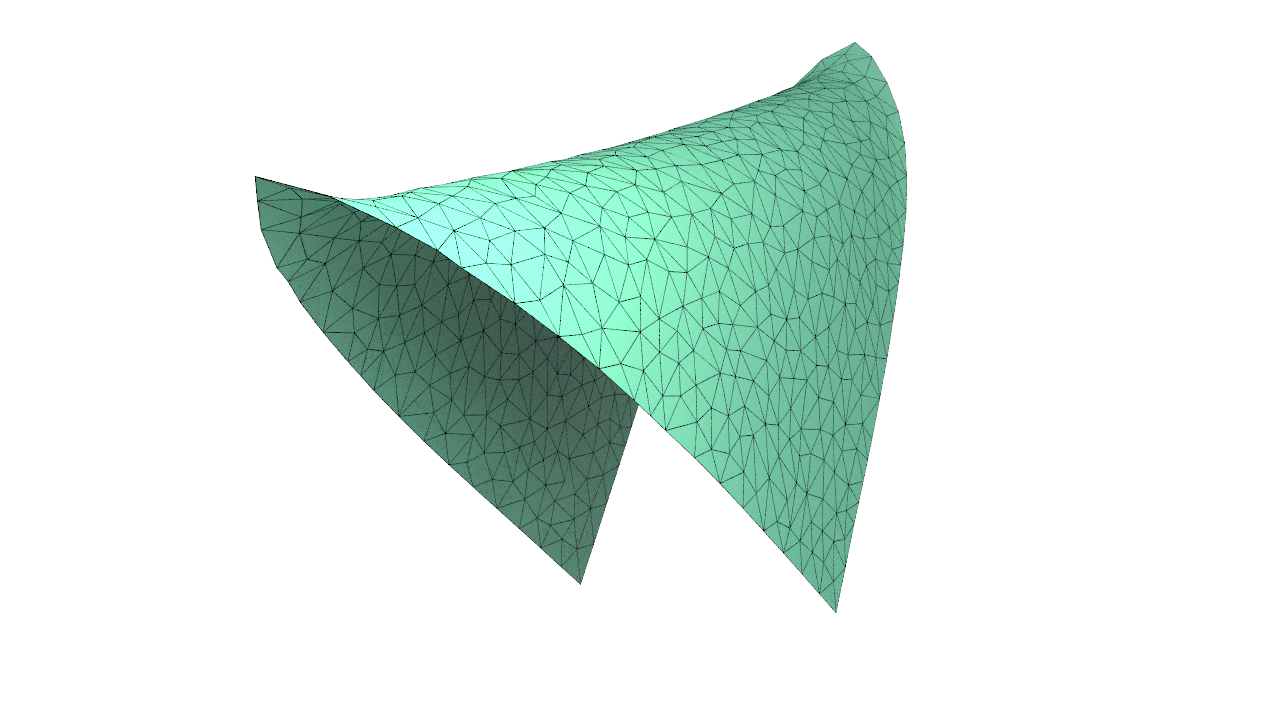}%
    \includegraphics[width=0.24\textwidth]{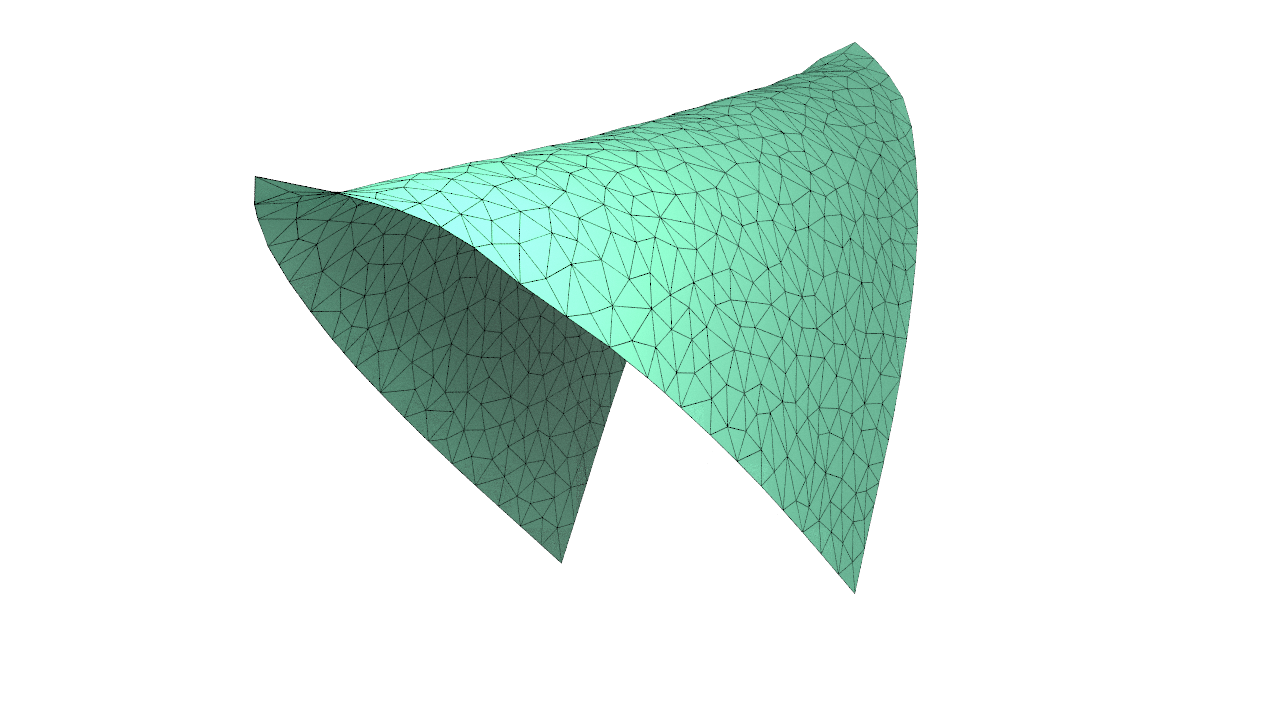}%
    \includegraphics[width=0.24\textwidth]{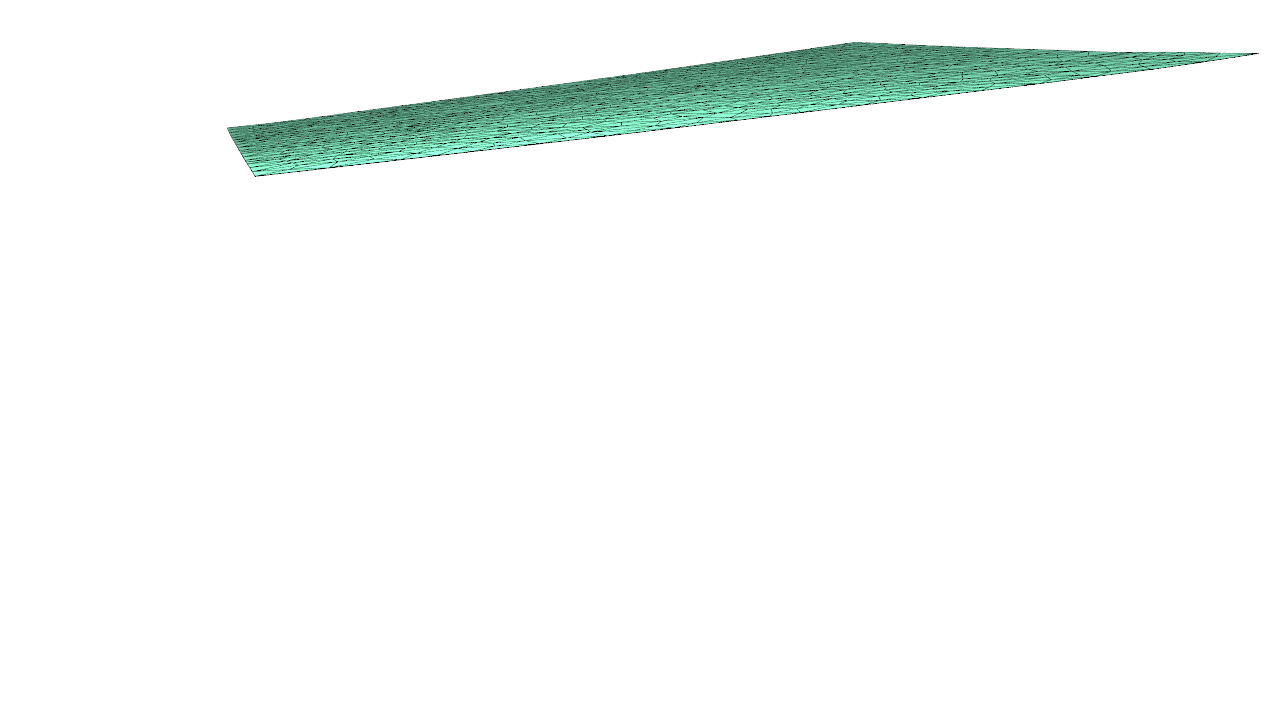}%
    \includegraphics[width=0.24\textwidth]{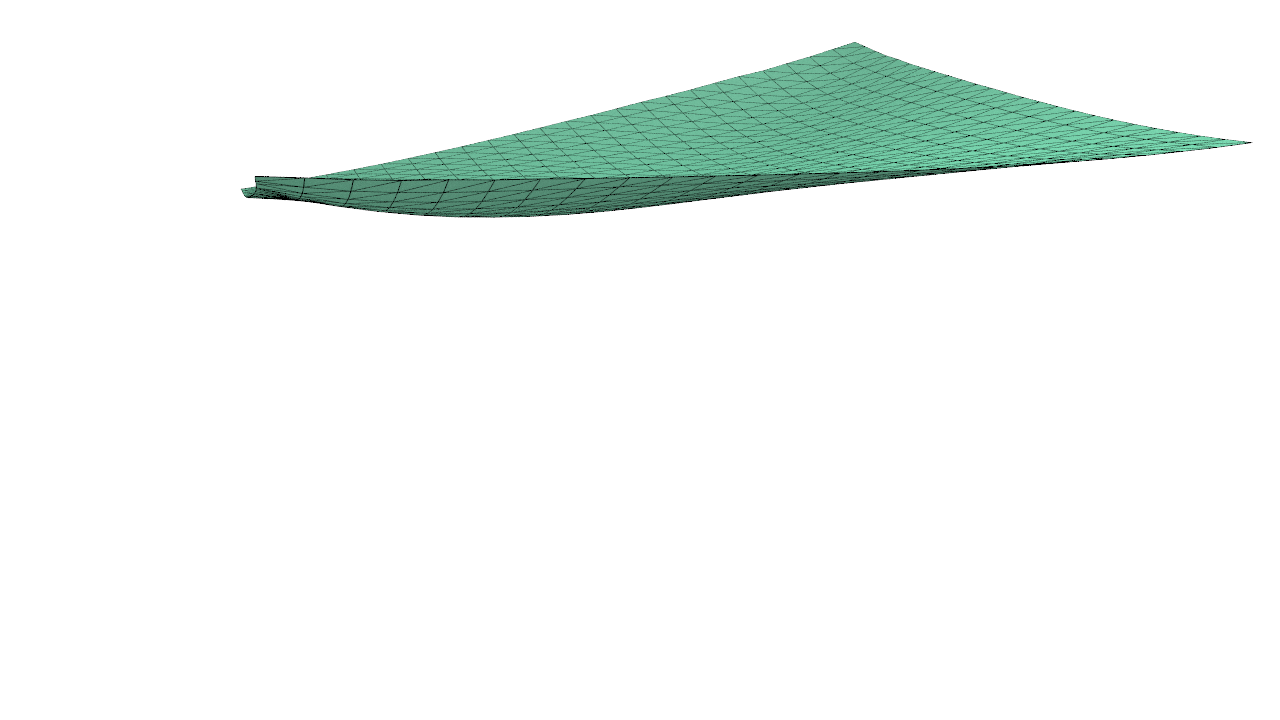}\\%
    \includegraphics[width=0.24\textwidth]{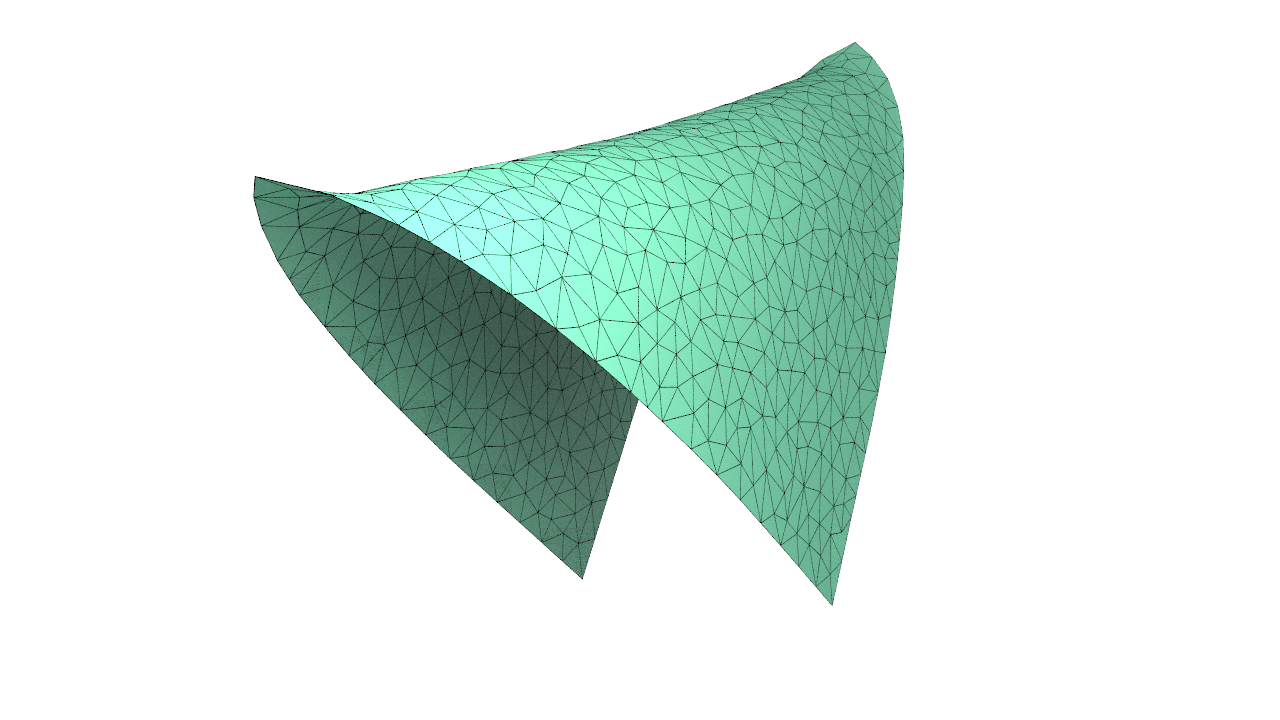}%
    \includegraphics[width=0.24\textwidth]{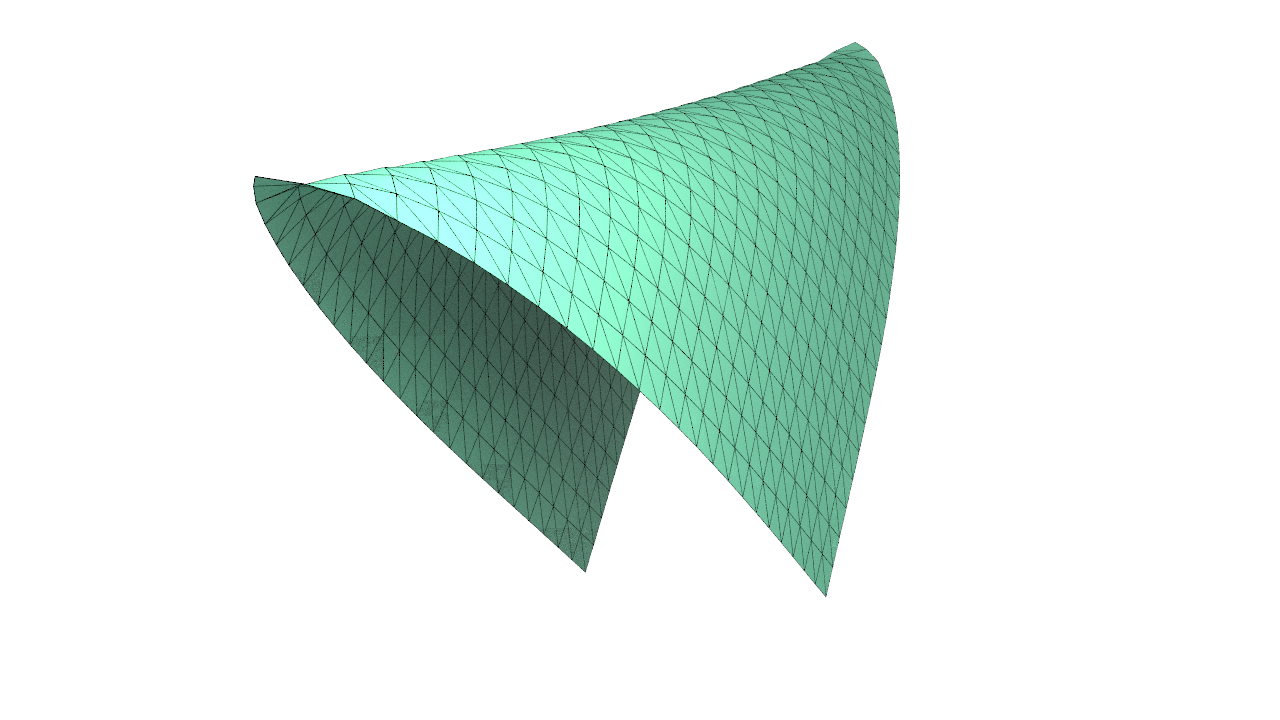} 
    \includegraphics[width=0.24\textwidth]{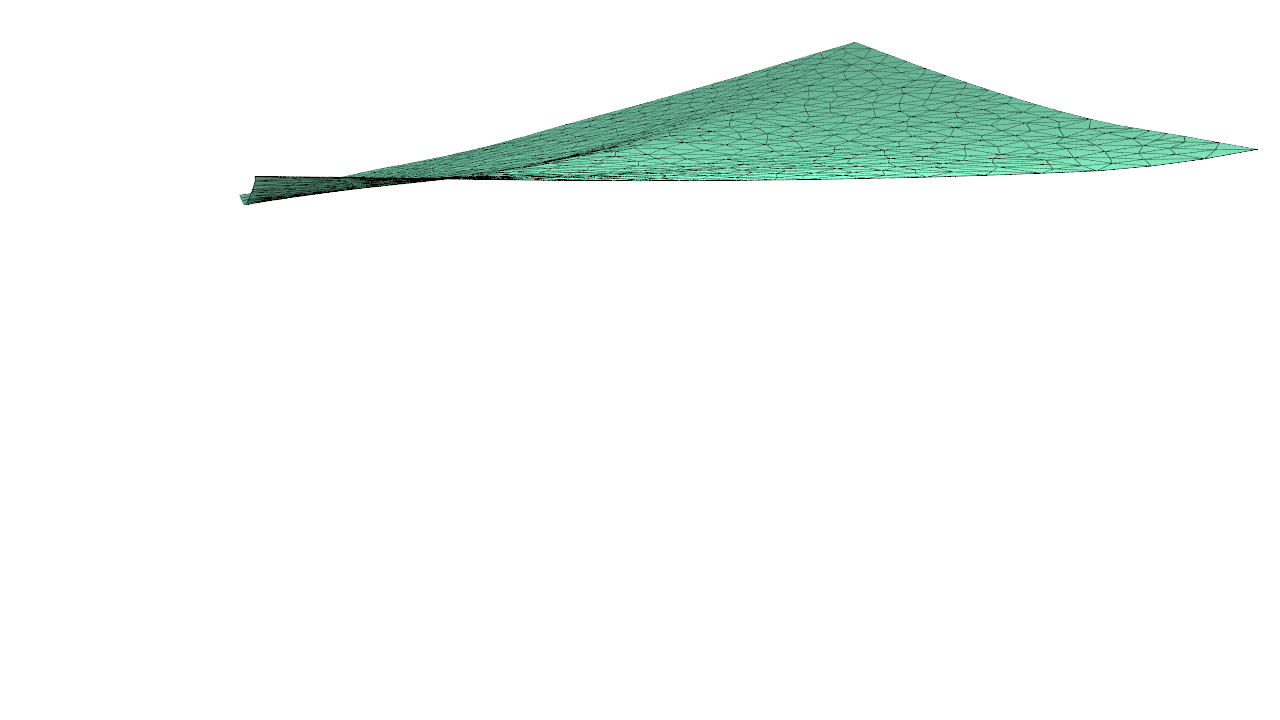}%
    \includegraphics[width=0.24\textwidth]{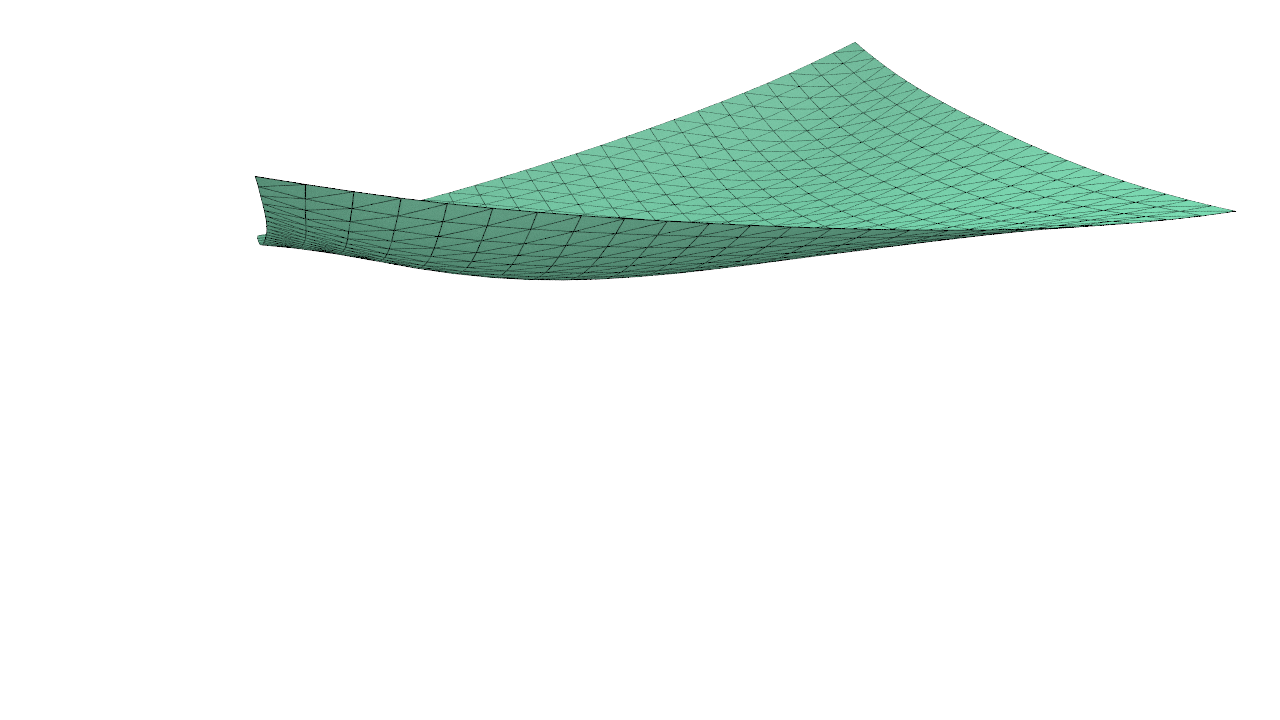}\\%
    \includegraphics[width=0.24\textwidth]{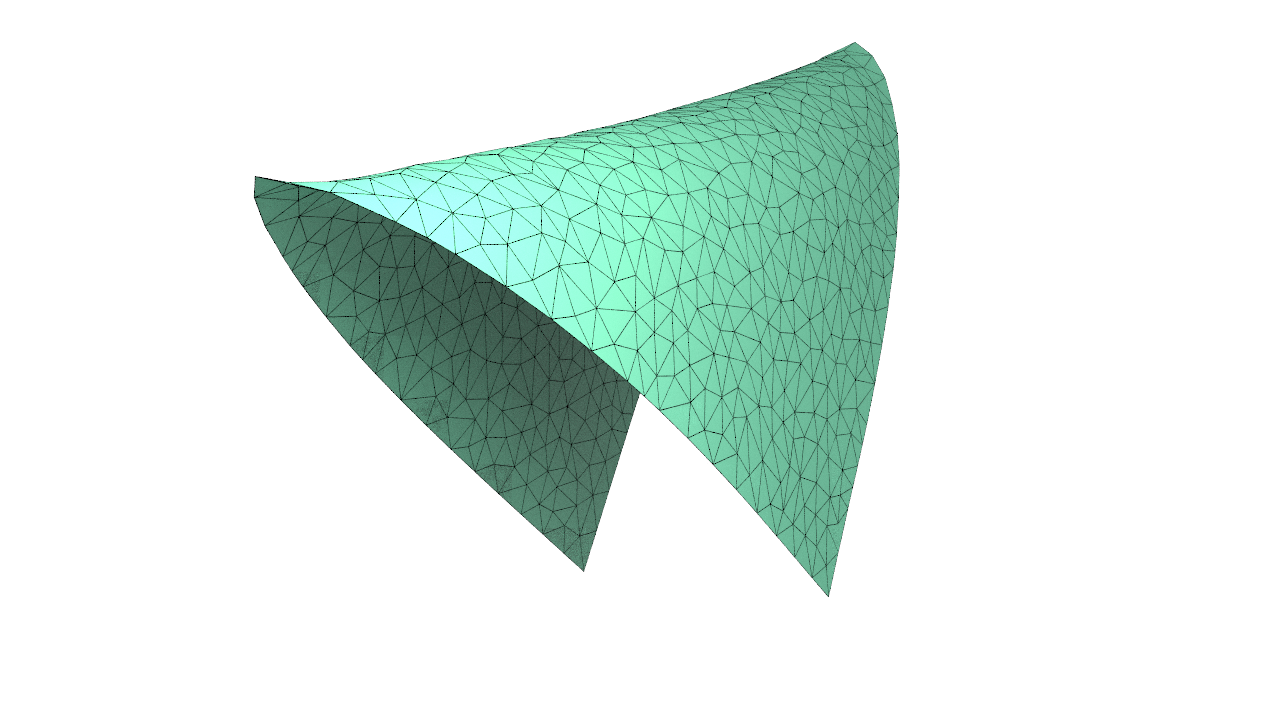}%
    \includegraphics[width=0.24\textwidth]{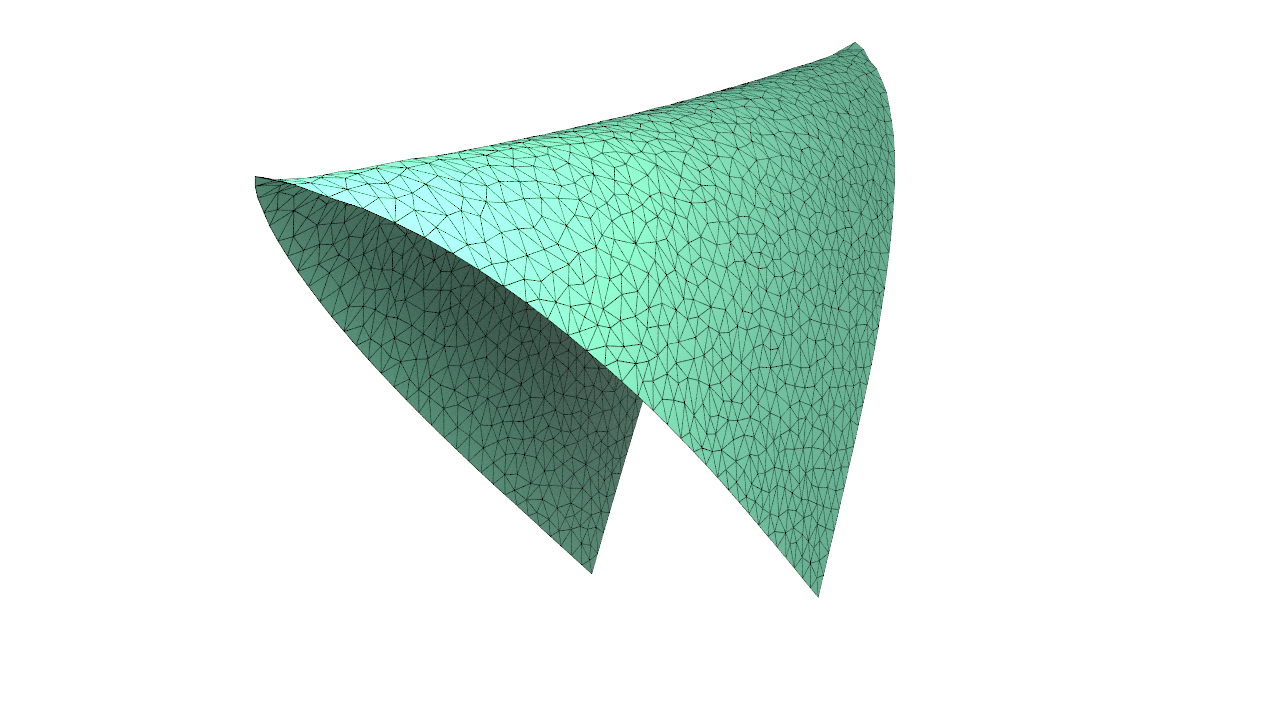}%
    \includegraphics[width=0.24\textwidth]{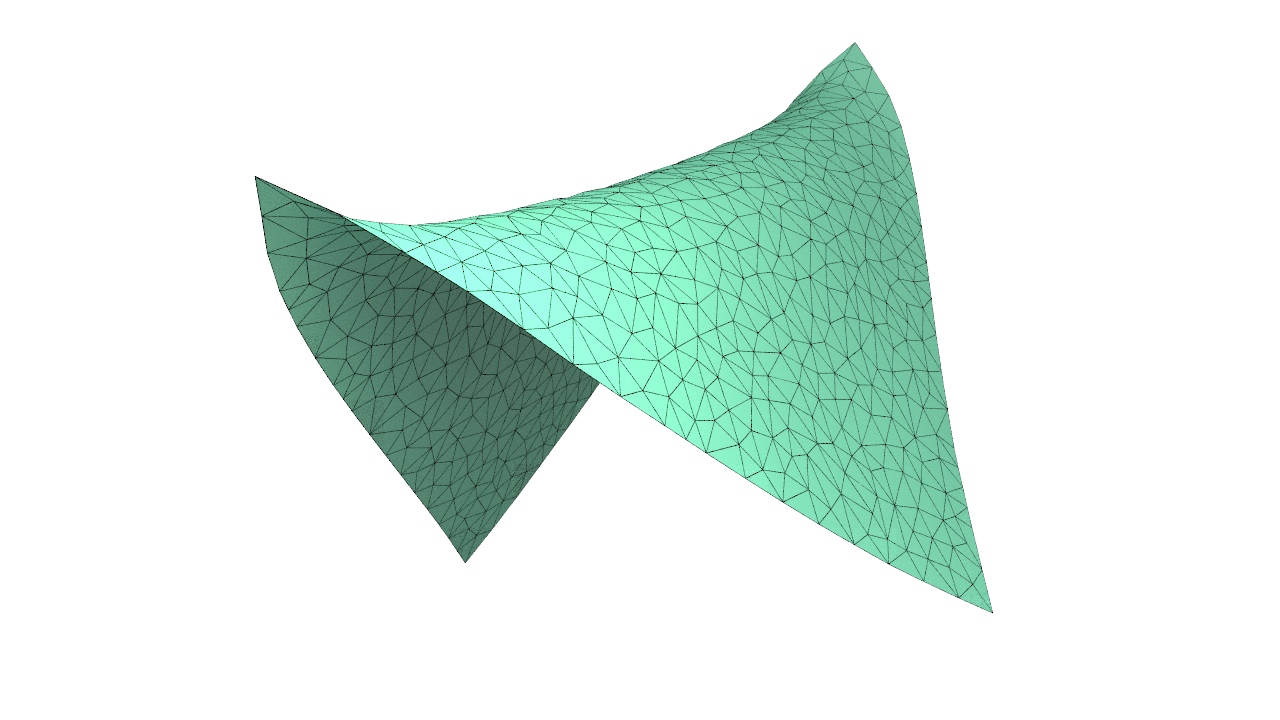}%
    \includegraphics[width=0.24\textwidth]{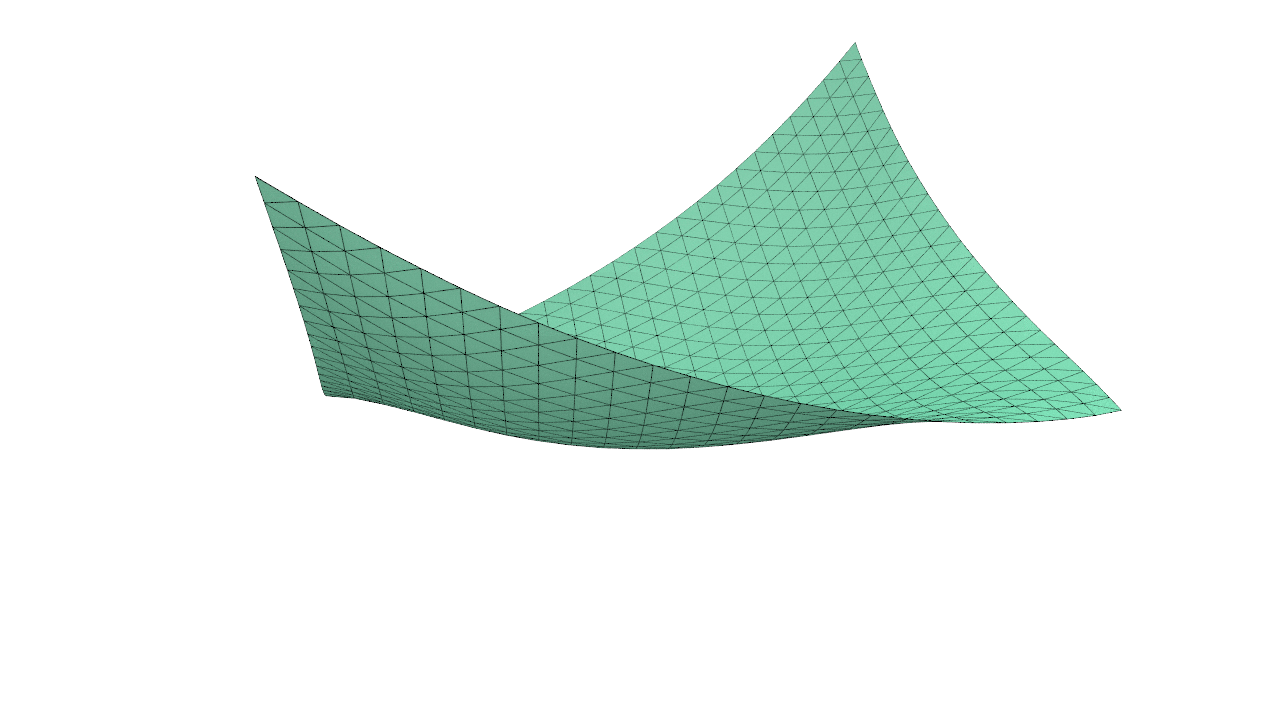}%
    \caption{Locking test on a square piece of cloth pinned at the corner. \emph{Left column}: simulation using our isometry constraints, with constraints tolerance $0.1$, $0.01$, and $0.001$ from top to bottom. \emph{Second column}: sampling with twice the neighborhood size, on a regular grid, and on a finer mesh. All these simulations have similar curvature and do not lock. \emph{Third column}: simulations using discrete elastic shells~\cite{Grinspun2006} for the same bending stiffness as our simulations, but varying stretching stiffnesses, decreasing from top to bottom (Young's moduli: $10^4$, $100$, and $1$ MPa). Notice there is no free lunch between excessive in-plane strain, and membrane locking. \emph{Last column}: Simulation using isometry constraints on all edges of a quad mesh with varying diagonal spring stiffness, decreasing from top to bottom~\cite{GHFBG07}.}
    \label{fig:pinnedcloth}
\end{figure*}
\section{Meshless Kinematics} \label{sec:kinematics}
In this section, we describe the meshless kinematics we use, and our notation. The heart of our method will be in Section~\ref{sec:constraints}, where we formulate isometry constraints on neighborhoods of the material.
We assume we are given a surface $\mathcal{S}$ whose rest state is flat. We adopt a standard meshless discretization of the plate, and sample $S$ at $N$ points $\mathcal{P}$ of $\mathcal{S}$, and associate to each point a neighborhood $\mathcal{N}_i \subset \mathcal{P}$ of other sample points. We require that $|\mathcal{N}_i| \geq 2$; the neighboring points can be chosen based on a threshold distance away from the sample points on $\mathcal{S}$, graph distance on a user-provided input triangle mesh, etc. 

We assume that locally, each neighborhood $\mathcal{N}_i$ can be isometrically parameterized by a region $\Omega_i$ of the plane (this parameterization might be given, in the case of cloth sewing patterns, or precomputed from $S$ by plane-fitting). We will write $X^i_j \in \mathbb{R}^2$ for the position of sample point $j \in \mathbb{N}_i$ on $\Omega_i$; note that a sample point likely belongs to multiple neighborhoods and might have different material coordinates $X^i_j$ in each such neighborhood. Let $Y_i\in\mathbb{R}^3$ be the embedded position of sample point $i$ in 3D. The set of $Y_i$ are then the degrees of freedom of the simulation. Finally we associate to each neighborhood a lumped mass $m_i$ (based on a given material density and the barycentric area of its associated sample point). 
\section{Isometry Constraints} \label{sec:constraints}

\begin{figure}[ht]
\includegraphics[width=0.49\columnwidth]{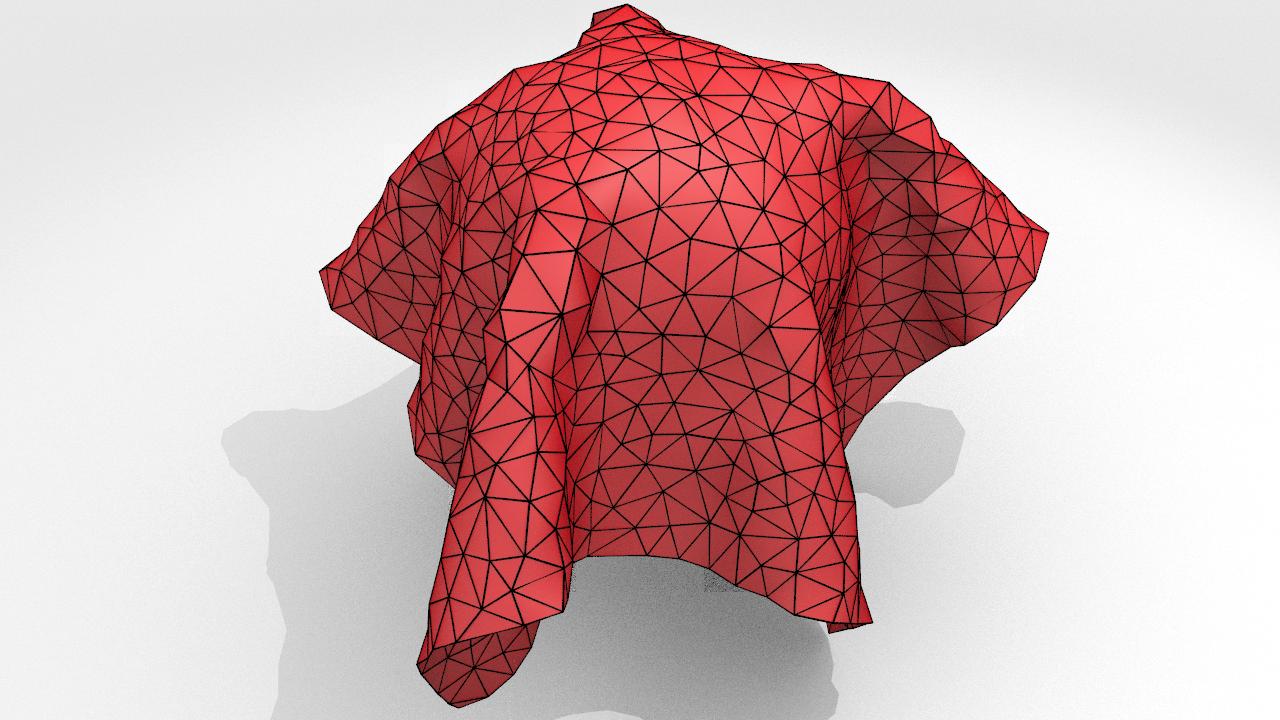}\,%
\includegraphics[width=0.49\columnwidth]{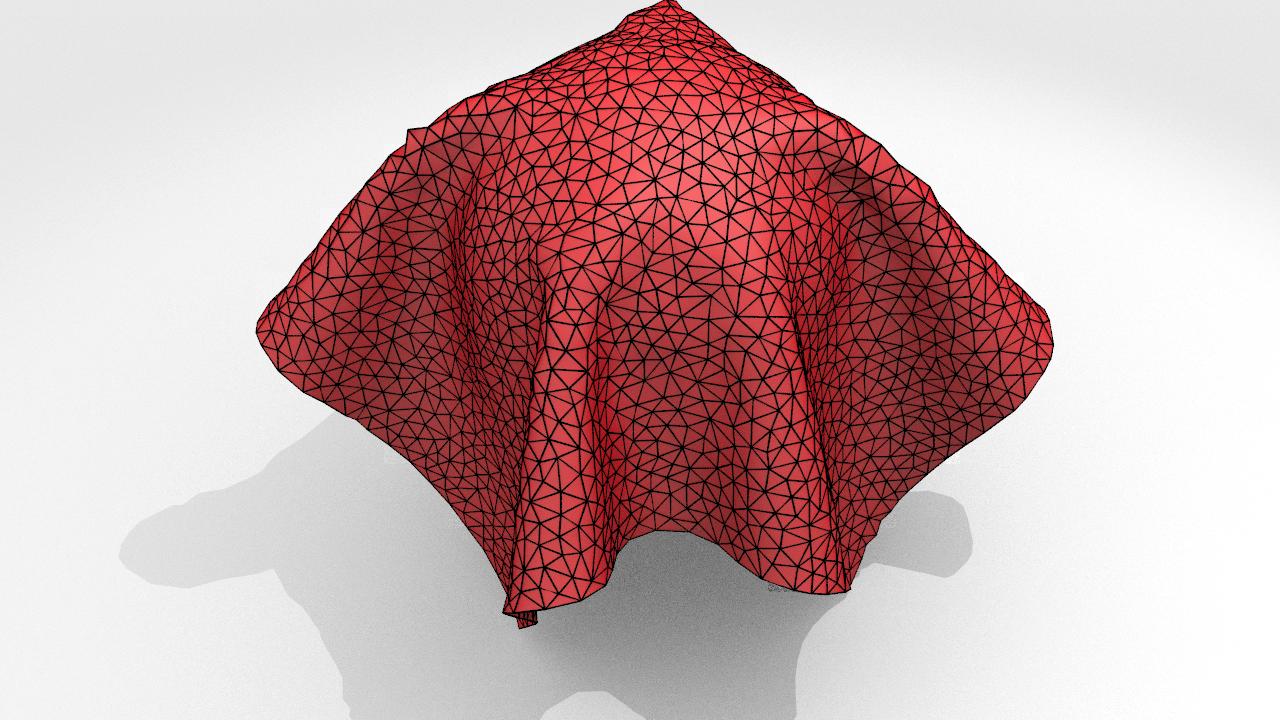}\\%
\includegraphics[width=0.49\columnwidth]{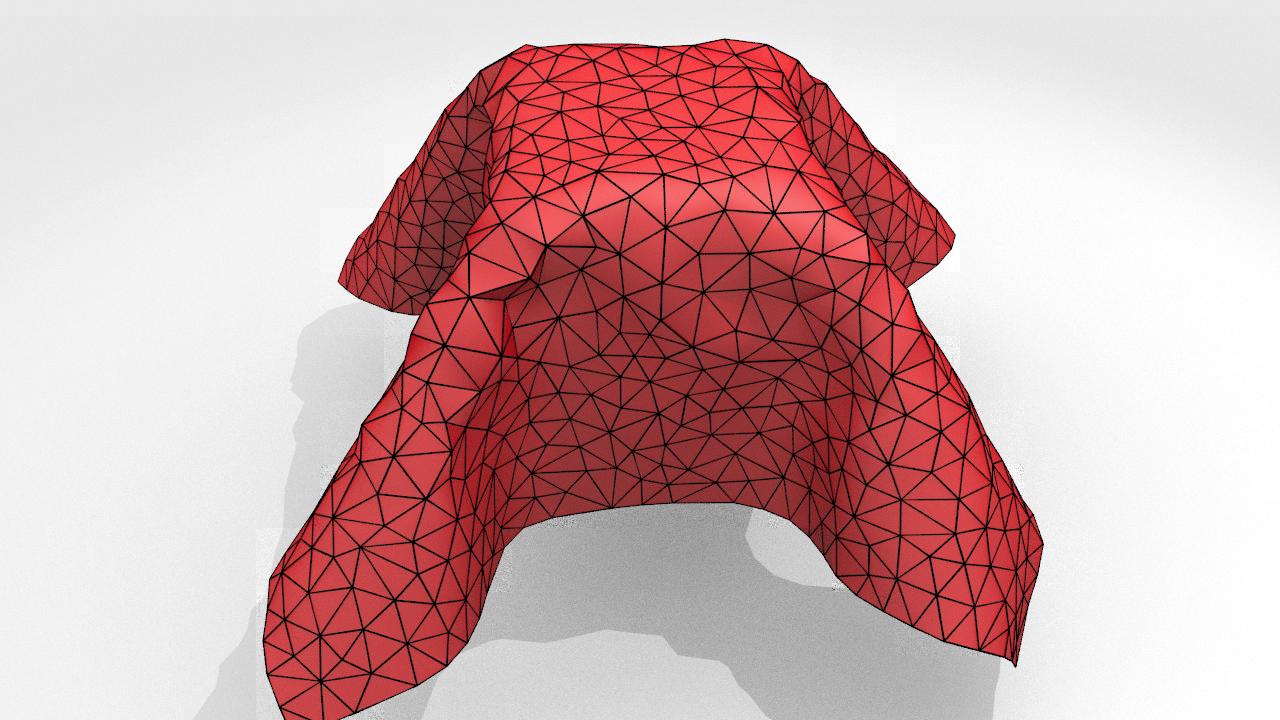}\,%
\includegraphics[width=0.49\columnwidth]{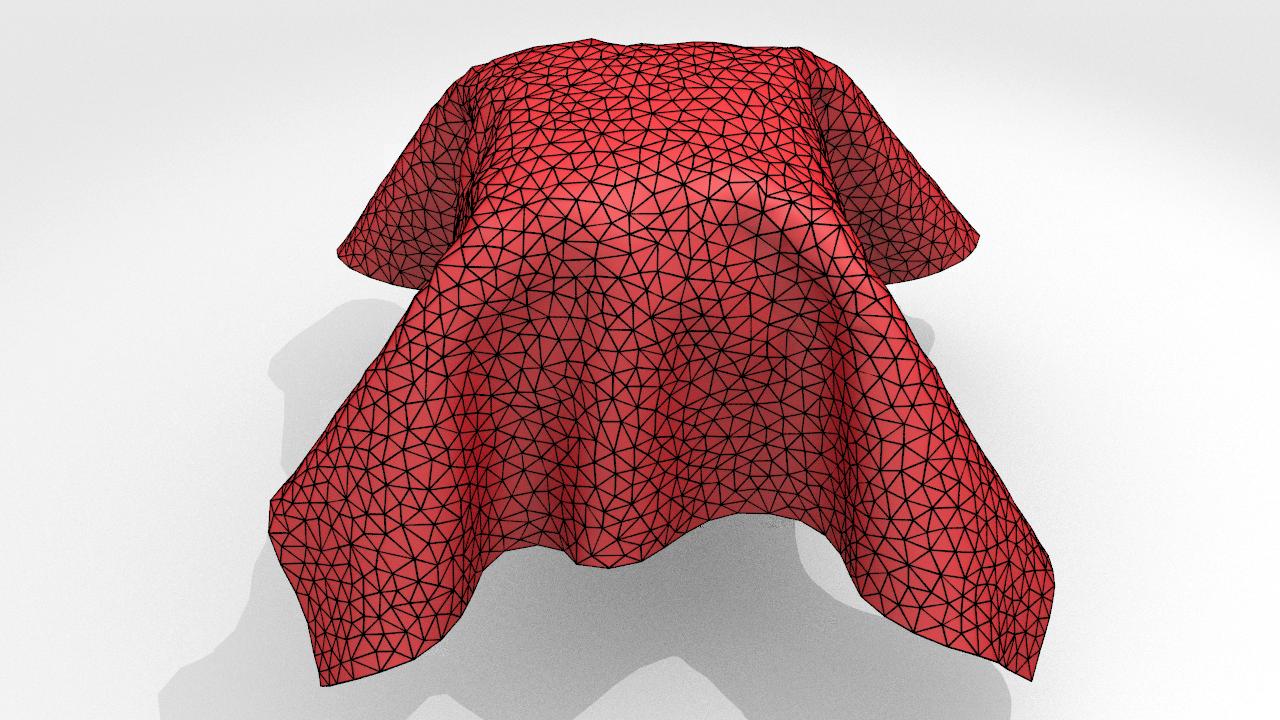}%
\caption{Draping experiments using coarse and fine meshes (left and right columns, respectively). Our isometry constraints yields consistent results even for very coarse discretizations of the cloth.}
\label{fig:draping}
\end{figure}

\paragraph{Motivation} The most obvious way to enforce isometry of a surface is to discretize it as a triangle mesh, and constrain each edge length to remain constant. However, such constraints are doomed to lock. Consider for instance that for a mesh with $|V|$ vertices and $|E|$ edges, there are $3|V|$ kinematic degrees of freedom and $|E|$ constraints, and yet from the Euler characteristic formula, $|E| \approx 3|V|$; the edge-based isometry constraints are therefore expected to remove almost all of the cloth's degrees of freedom. Contrast this situation with the smooth setting, where the cloth can deform into a rich variety of developable surfaces. Note that replacing the triangle mesh with quadrilaterals does not in any way solve the problem: one can then enforce isometry for only the quadrilateral edges (as in Goldenthal et al.~\shortcite{GHFBG07}), which allows non-isometric shear and stretching in the diagonal direction (see Figure~\ref{fig:pinnedcloth}), or also constrain the lengths of the diagonals, which essentially triangulates the mesh.

We propose instead to enforce isometry on the \emph{vertices}: we will use moving least-squares to formulate a strain on each of the $|V|$ neighborhoods $\mathcal{N}_i$, and constraint the principle strains to be zero. We will then have only $2|V|$, rather than $|E|$, isometry constraints, leaving $|V|$ leftover degrees of freedom for isometric bending modes.

\paragraph{Constraint Formulation} Near a sample point $i$, the $3\times 2$ deformation gradient $F$ of the thin plate, with respect to the local parameterization of $\mathcal{N}_i$, must satisfy
\begin{equation} 
FX^i_j \approx Y_j - Y_i \quad \forall j \in \mathcal{N}_i. \label{eq:strain}
\end{equation}
Of course, for points $i$ with more than two neighbors, this relation is overconstrained; we instead work with an \emph{averaged} deformation gradient $F_i$ in $\mathcal{N}_i$, based on satisfying Equation~\ref{eq:strain} in the moving least-squares sense,
$$F_i = \argmin_{F} \sum_{j\in\mathcal{N}_i} m_{j} \left\|FX^i_j - (Y_j-Y_i)\right\|^2.$$ 
The averaged deformation gradient $F_i$ can then be expressed in closed form as
$$F_i = \bY \bW \bX^T (\bX \bW \bX^T)^{-1},$$ 
where each column of $\bX_{2\times |\mathcal{N}_i|}$ is one $X^i_j$, each column of $\bY_{3\times |\mathcal{N}_i}$ is one current displacement $Y_j-Y_i$, and $\bW_{|\mathcal{N}_i|\times |\mathcal{N}_i|} = \operatorname{diag}(m_{j}).$

\paragraph{Isometry} Given current deformation gradient $F_i$ for a neighborhood of point $i$, we can formulate the strain tensor, 
$$\varepsilon_i = (F_i^TF_i - I),$$ 
for $I$ is the identity matrix. Typically for dynamics we would then derive forces by applying a constitutive law to this strain; since we instead want to enforce inextensibilty as a hard constraint, we need to write down constraint functions specifying vanishing of the strain. There are several sets of constraints we could choose; unfortunately, all are nonlinear. We use the pair
\begin{align}
\begin{split}
g^{\mathrm{tr}}_i &= \operatorname{tr}\left(F_i^TF_i\right) = 0 \\
g^{\mathrm{det}}_i &= \det\left(F_i^TF_i\right) = 0,
\end{split}
\label{eq:constraints}
\end{align}
Since the zero matrix is the only symmetric matrix with both eigenvalues zero, these constraints are equivalent to $\varepsilon_i=\bm{0}.$ The gradients of these constraints, in local coordinates with respect to a variation $\delta \bY$ of $\bY$, are given by
\begin{align*}
    \nabla g^{\mathrm{tr}}_i \cdot \delta \bY &= 2F(\bX \bW \bX^T)^{-1}\bX\bW : \delta \bY\\
    \nabla g^{\mathrm{det}}_i \cdot \delta \bY &= 2F(F^TF)^{\mathrm{adj}}(\bX \bW\bX^T)^{-1}\bX \bW : \delta \bY,
\end{align*}
where
$$\left[\begin{array}{cc}a & b\\c & d\end{array}\right]^{\mathrm{adj}} = \left[\begin{array}{cc}d & -b\\-c & a\end{array}\right].$$
Notice in particular that both constraints have zero as a regular value: neither gradient vanishes when $\varepsilon_i=0$, a crucial requirement for the numerical stability of techniques like Fast Projections that are based on the method of Lagrange multipliers. Intuitively, the trace of the strain tensor corresponds to the sum of the squared principle stretches, which is in a sense the averaged squared distance to the neighboring points. The determinant of the strain tensor is the product of the squared principle stretches and measures a notion of squared neighborhood area. Both constraints together imply that the stress tensor, averaged in each neighborhood, is the zero matrix, equivalent to isometry of the neighborhood.

\begin{figure}[!t]
\includegraphics[width=\columnwidth]{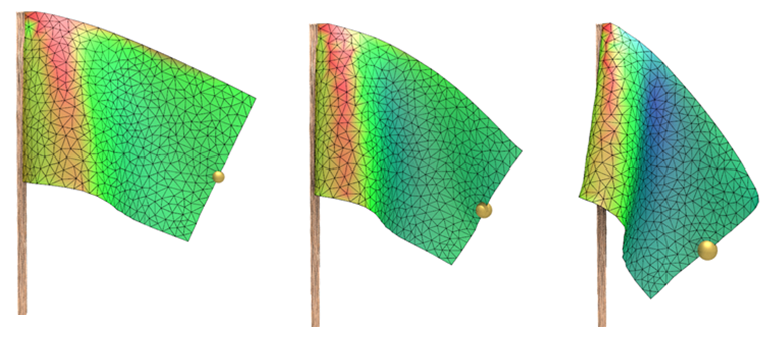}%
\caption{We test the behavior of our constraints on a shear-dominated scenario. Warmer color indicates an increase in extrinsic distance from the given point to the top left corner of the flag. The maximum increase in distance is under 1\%.}
\label{fig:shear}
\end{figure}

We mention in passing some alternative sets of constraints we tried, and abandoned: perhaps the most obvious is to directly constrain each entry of $\varepsilon_i$, which amounts to three constraints per neighborhood since strain is symmetric. However, this formulation bloats the set of constraints by 50\%, while introducing rank deficiency in the constraint gradients, leading to poor performance. One might also consider constraining the trace and determinant of the strain tensor, rather than of $F_i^TF_i$---the problem with this approach is that the function $\det(\varepsilon_i)$ has a critical point at the strain-free state, which as mentioned above makes the constraint unsuitable for projection.

\paragraph{Relation of Locking to Constraint Jacobian Rank} We stress that the locking phenomena described in the introduction, and the first paragraph of this section, are entirely unrelated to linear dependence or independence of the constraint gradients. Obviously, unsatisfiable constraints cannot be enforced. But note that constraining the length of each edge in a triangle mesh locks the mesh, despite these constraints being linearly independent. Linearly independent constraint gradients are insufficient to prevent locking, nor will redundant (dependent) constraint gradients cause locking.

Our constraints are always satisfiable (assuming that the initial configuration is an isometric embedding of the plate) but the constraint gradients are not guaranteed to be independent: for instance, when $\varepsilon_i=0$, the trace and determinant constraints have identical gradient. This linear dependence is expected, from the geometry of bending: a flat neighborhood has two isometric deformation modes, since it can bend in any direction, but once the neighborhood begins to bend in one particular direction and symmetry is broken, only one isometric mode remains.

\section{Time Integration} \label{sec:integration}
We enforce the isometry constraints~\eqref{eq:constraints} at every time step using the method of Fast Projections~\cite{GHFBG07}, with implicit time integration of bending and external forces. Any bending model can be used, though we note that for materials which are rest flat (including cloth, paper, etc), the simple and efficient quadratic bending model~\cite{Bergou2006} is quite attractive, since isometry of the material is precisely the assumption required for validity of the model:
$$E_{\mathrm{bend}} = \frac{k}{2} \|L\bY\|^2_{M^{-1}},$$
where $M$ is the system mass matrix, $L$ the Laplace-Beltrami operator (computed from a meshless discretization~\cite{Petronetto2013}, or from a triangulation of the input surface), and $k$ is a bending stiffness proportional to the Young's modulus and cubed thickness of the material. Note that the quadratic bending energy, true to its name, has constant Hessian which can be prefactored, so that the performance bottleneck of each time integration step is enforcing the isometry constraints.

In most of our experiments, we used a time step size of $10^{-3}$ seconds and a constraint tolerance of 0.1; Fast Projections typically converges in one to three iterations. However, Fast Projections does not perform a true projection onto the constraint manifold, and comes with no guarantees; we found that regularizing each projection by the geometric stiffness matrix of the constraints~\cite{Tournier2015} improves performance. In cases where Fast Projections fails to decrease the constraint residual after a few iterations, we switch to true projection using the augmented Lagrangian method; in practice this fallback was only needed when high-energy impacts cause large violations in the constraints within a single time step.
\section{Results} \label{sec:results}

\paragraph{Locking Tests}

We perform a sequence of tests demonstrating that the constraints~\eqref{eq:constraints} neither lock nor exhibit spurious modes. In Figure~\ref{fig:pinnedcloth}, we pin a square sheet of cloth at two corners, and let the sheet relax under gravity. We simulate this experiment using our isometry constraints for different constraint tolerances, for different resolutions of sampling points, and for both ordered and disordered sampling of the cloth. Our simulations consistently give comparable results. Similarly, for tests that drape cloth on rigid bodies (Figure~\ref{fig:draping}), our method produces qualitatively similar results, without locking, even for very coarse discretizations of the cloth. Statistics for these and other experiments described in this section can be found in Table~\ref{tab:numbers}.

\paragraph{Effect of Parameters} As we decrease the constraint tolerance as shown in the 
left column
in Figure~\ref{fig:pinnedcloth}, the sagging in the middle of the cloth decreases as expected, since the isometry constraints enforce that the distance between the two pinned corners is identical to that of the flat rest shape. 

\paragraph{Comparison to Related Work} We compare to force-based shell methods~\cite{Grinspun2006}, using the same bending stiffness, on the 
third column
of Figure~\ref{fig:pinnedcloth}, where enforcing isometry via high stiffness locks the cloth. Lastly, the same experiment was performed using the constraints proposed by~\citet{GHFBG07}. Due to the fact that the edge length constraints on a quad method do not control the length of each quad diagonal, the cloth sags significantly.   

We also perform a test where the cloth deformation is driven by shear. We pin one side of a piece of cloth, and attach a heavy mass to a point on the other. The cloth stretches taut along diagonal lines of tension, but does not lock, and in contrast to the Goldenthal et al.~constraints on the principal strain along the axes of a quadrilateral grid, obeys the isometry constraints, which we verify by plotting the change in extrinsic distance from the top left corner to every other points in the material; see Figure~\ref{fig:shear}.

\begin{figure}[ht]
\includegraphics[width=\columnwidth]{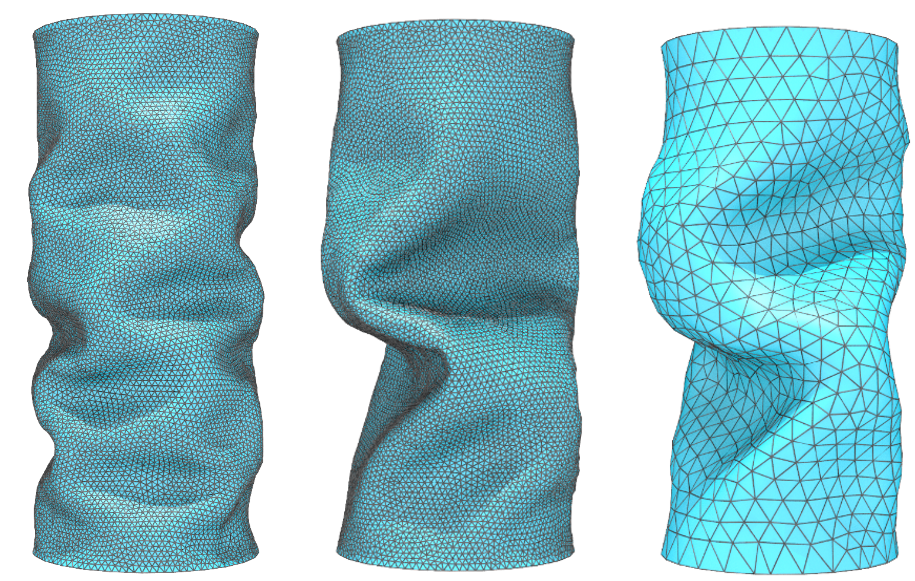}%
\caption{Frames of cylinder crumpling using our method, for a fine (\emph{left and middle}) and coarse (\emph{right}) simulation. Only the fine simulation resolves the fine-scale pattern but both simulations settle to similar static states.}
\label{fig:cylinder}
\end{figure}

\paragraph{Crushed Cylinder} Our method, unlike thin plate formulations based on tension field theory, works equally well in problems involving pure compression, such as the crumpling of an elastic cylinder. We replicate the cascade of diamond buckling patterns observed by \citet{Martin2010}, and although only our high-resolution simulation can resolve the fine-scale pattern at the beginning of the simulation, it correctly reproduces the highly-buckled final state.

\begin{figure}[h]
\includegraphics[width=\columnwidth]{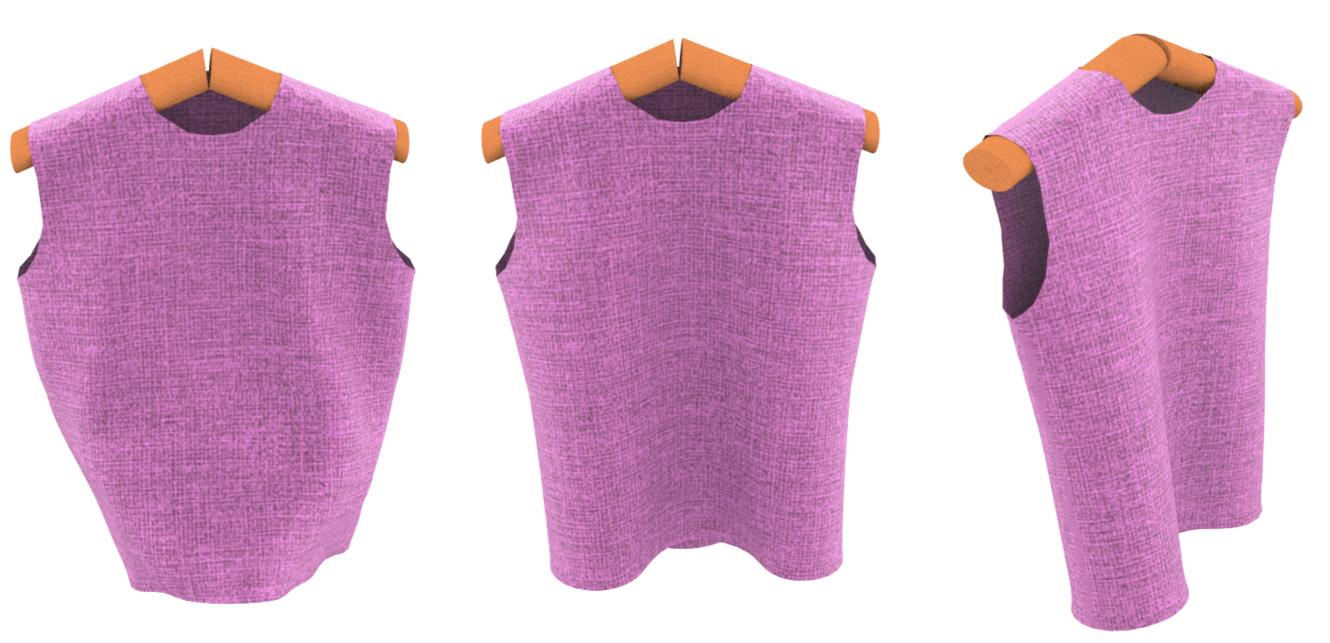}%
\caption{Dynamics of a spinning tank top.}
\label{fig:tank}
\end{figure}

\paragraph{Spinning tank top} We show the dynamics of a tank top draping over a spinning hanger in Figure~\ref{fig:tank}.  Our method resolves the wrinkles generated during the motion.

\begin{figure}[h]
\includegraphics[width=\columnwidth]{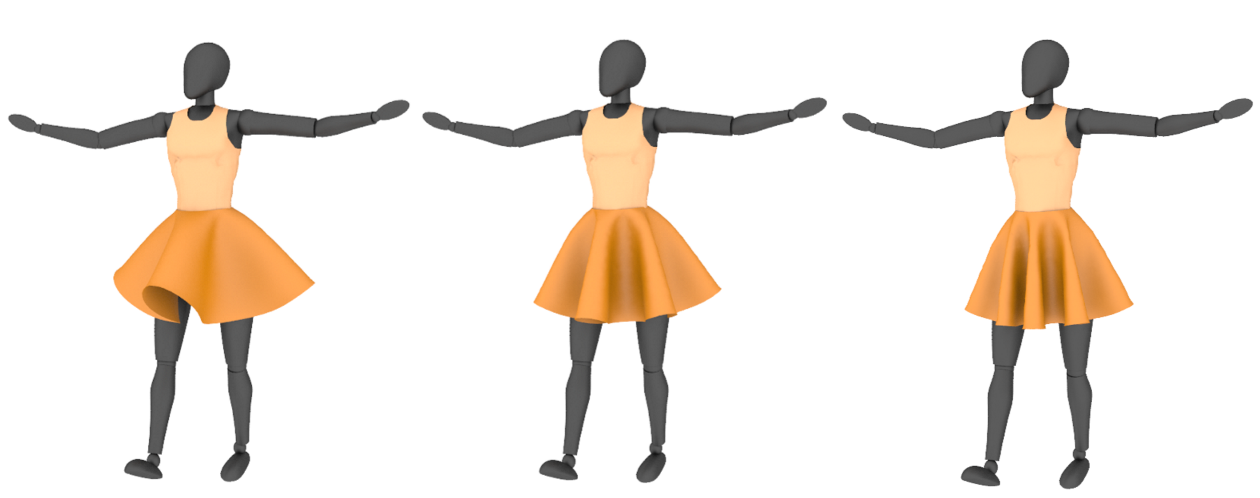}%
\caption{Our method generates draped skirt shapes with different wrinkle patterns while preserving the isometry.}
\label{fig:skirt}
\end{figure}

\paragraph{Bending Stiffness} One important feature of our method
is that the bending stiffness of the simulated cloth can be controlled, while preserving isometry. For most thin shell models, extensive tuning of the hyper-parameters is often required to achieve the balance between avoiding locking and maintaining isometry of the material. As shown in Figure~\ref{fig:skirt}, just by adjusting the bending stiffness, our method produces cloth with different wrinkling patterns without stretching.

\begin{figure*}[ht]
\includegraphics[width=0.33\textwidth]{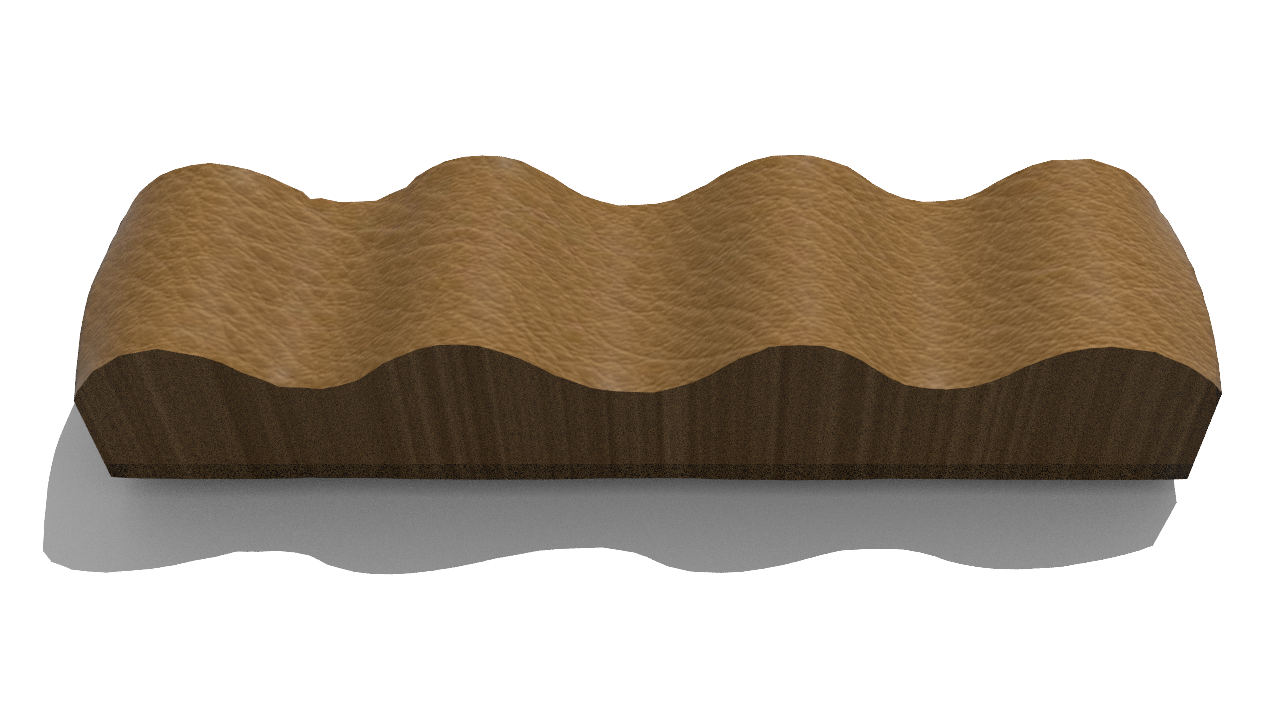}%
\includegraphics[width=0.33\textwidth]{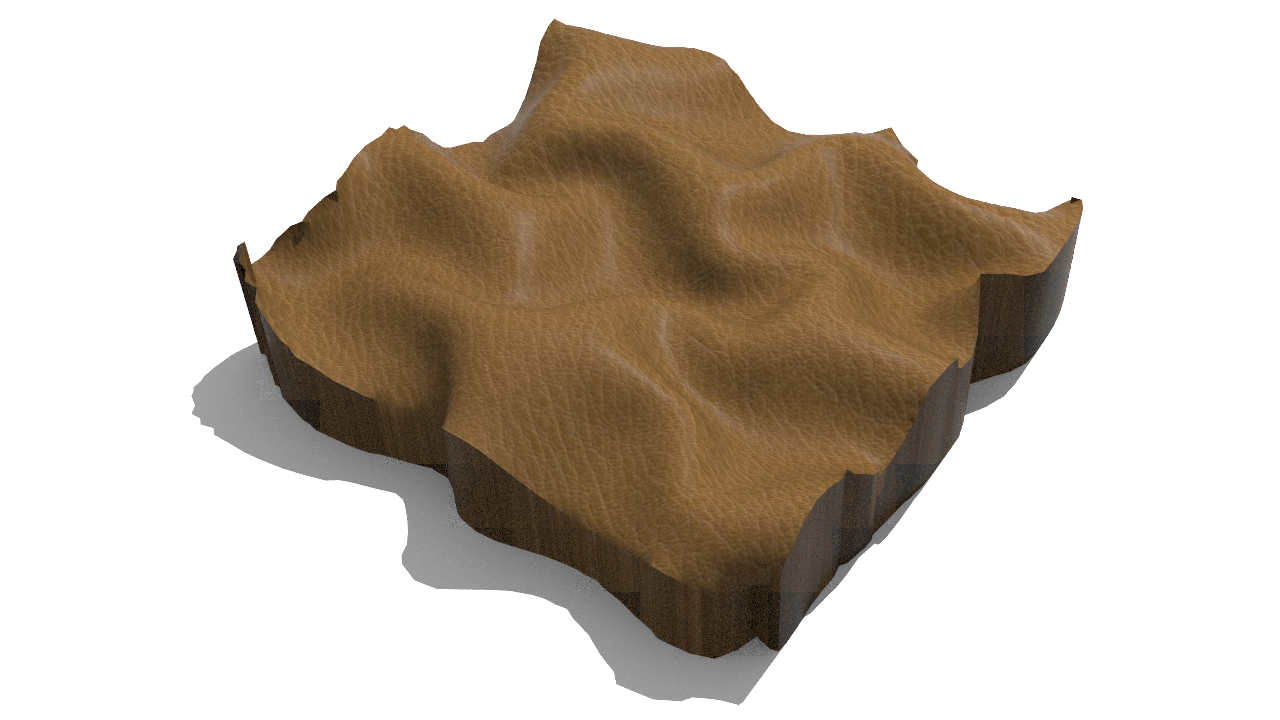}%
\includegraphics[width=0.33\textwidth]{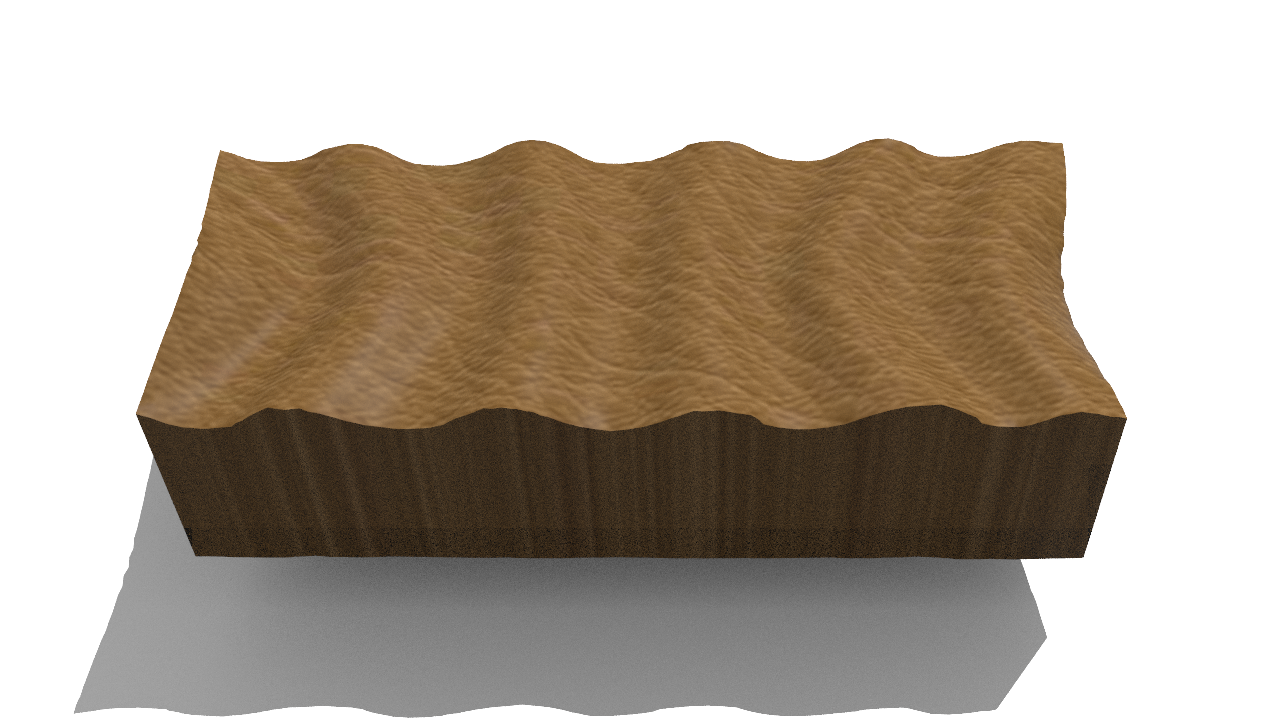}%
\caption{Simulations of skin coupled to a volumetric substrate. Even for coarse discretizations, our method reproduces the correct undulating or 3D-folded wrinkling pattern for compression in one or two directions, and pinching.}
\label{fig:skin}
\end{figure*}

\paragraph{Skin Wrinkling} Several papers~\cite{Rmillard2013EmbeddedTS,Li:2014:MSS:2668064.2668089} describe how to simulate wrinkling of skin via one-way coupling of thin plate simulations to a volumetric elastic substrate. The coupling is enforced via a sparse set of average-position constraints sampled on the skin surface. This setup is ideal for our method, since (i) the behavior of the upper skin layers is governed by bending and by the coupling constraints, so that replacing in-plane strain of the skin with isometry of these fine layers is justified; (ii) the coupling constraints can be trivially incorporated by including them as extra terms during Fast Projections. Figure~\ref{fig:skin} shows the behavior of our simulation when the substrate is compressed in one and two directions.

\begin{table*}[ht]
\caption{Parameters used in each experiment.}
\centering
\begin{tabular}{*{9}{|c}|} 
\hline
   & Fig. ~\ref{fig:pinnedcloth} coarse  & Fig. ~\ref{fig:pinnedcloth} regular & Fig. ~\ref{fig:pinnedcloth} fine & Fig. ~\ref{fig:draping} coarse & Fig. ~\ref{fig:draping} fine & cylinder(coarse) & cylinder (fine) & tank top\\
 \hline
 verts & 662  & 625 & 1656 & 845 & 2716 & 979 & 15039 & 3450 \\
 \hline
 tolerance & 0.01 & 0.01 & 0.01 & 0.01 & 0.01 & 0.001 & 0.001 & 0.01 \\ 
 \hline
\end{tabular}
\label{tab:numbers}
\end{table*}


\bibliographystyle{ACM-Reference-Format}
\bibliography{InextensibleShell}



\end{document}